\documentclass[lettersize,journal]{IEEEtran}
\usepackage{amsmath,amsfonts,amssymb}
\usepackage{algorithmic}
\usepackage{algorithm}
\usepackage{array}
\usepackage[caption=false,font=normalsize,labelfont=sf,textfont=sf]{subfig}
\usepackage{textcomp}
\usepackage{stfloats}
\usepackage{xurl}
\usepackage{verbatim}
\usepackage{graphicx}
\usepackage{hyperref}
\usepackage{cleveref}
\usepackage{xcolor}
\usepackage{tabularx}
\usepackage{siunitx}
\usepackage{booktabs}
\definecolor{load}{HTML}{5e7a2f}
\definecolor{NF1}{HTML}{AA4499}
\definecolor{NF2}{HTML}{882255}
\definecolor{NF3}{HTML}{CC6677}
\definecolor{slack}{HTML}{117733}
\definecolor{redTAGreg}{HTML}{ff0000}
\definecolor{blue_line}{HTML}{658cb2}
\definecolor{traj_turquoise}{HTML}{00e8d6}
\definecolor{traj_gray}{HTML}{828282}
\usepackage[style=ieee,doi=false,isbn=false,url=false]{biblatex}
\begin{document}


\title{Predicting Fault-Ride-Through Probability of Inverter-Dominated Power Grids using Machine Learning}


\author{\IEEEauthorblockN{Christian Nauck\IEEEauthorrefmark{1},
Anna Büttner\IEEEauthorrefmark{1},
Sebastian Liemann\IEEEauthorrefmark{2}, 
Frank Hellmann\IEEEauthorrefmark{1}, and
Michael Lindner\IEEEauthorrefmark{1}}
\\
\IEEEauthorblockA{\IEEEauthorrefmark{1}Potsdam Institute for Climate Impact Research, Telegrafenberg A31, 14473 Potsdam, Germany}
\\
\IEEEauthorblockA{\IEEEauthorrefmark{2}Institute of Energy Systems, Energy Efficiency and Energy Economics (ie3), Technical University of Dortmund, Dortmund, Germany}}





\maketitle
\begin{abstract}
    Due to the increasing share of renewables, the analysis of the dynamical behavior of power grids gains importance. Effective risk assessments necessitate the analysis of large number of fault scenarios. The computational costs inherent in dynamic simulations impose constraints on the number of configurations that can be analyzed. Machine Learning (ML) has proven to efficiently predict complex power grid properties. Hence, we analyze the potential of ML for predicting dynamic stability of future power grids with large shares of inverters. For this purpose, we generate a new dataset consisting of synthetic power grid models and perform dynamical simulations. As targets for the ML training, we calculate the fault-ride-through probability, which we define as the probability of staying within a ride-through curve after a fault at a bus has been cleared. Importantly, we demonstrate that ML models accurately predict the fault-ride-through probability of synthetic power grids. Finally, we also show that the ML models generalize to an IEEE-96 Test System, which emphasizes the potential of deploying ML methods to study probabilistic stability of power grids.
    
\end{abstract}


\section{Introduction}
    \IEEEPARstart{F}{uture} power grids will be operated by up to 100~\% renewable energy sources (RES), which poses challenges for power grid stability. RES-dominated grids are characterized by reduced inertia, more decentralized generation, and a more volatile production. These challenges increase the importance of analyzing and understanding the dynamical stability of power grids. RES are primarily linked to the power grid using power electronic inverters. There is an increasing need for grid-forming inverters \cite{ensoe_grid_forming_2020} that have the capability to stabilize RES-dominated grids without relying on conventional generation. Stability analyses that consider an individual grid-forming inverter connected to an infinite bus bar exist for almost all proposed control designs. Comprehensive network stability evaluations for grids dominated by novel grid forming actors are performed less often, with \cite{schifferConditionsStabilityDroopcontrolled2014} and \cite{colombinoGlobalPhaseVoltage2017} being notable exceptions.

    Dynamic simulations of models of real power grids are challenging with respect to appropriate modeling and the computational effort. Using dynamic simulations, to comprehensively assess stability in all relevant configurations, is currently not feasible. In grid operation, this limits the number of analyzable faults. As a consequence, critical scenarios could be missed and operation has to be relatively conservative, for example in line loading. During grid extension planning, dynamics are typically considered after an economic optimization is performed. Because of the increasing relevance of dynamic effects, the planing of future power grids should include a wider range of dynamic effects earlier, to possibly avoid the need for costly compensation measures.

    One approach to sidestep the need for detailed simulations is the implementation of Machine Learning (ML). In the context of power grids, ML has been used for various applications \cite{xieReviewMachineLearning2020,kumbharComprehensiveReviewMachine2021,liaoReviewGraphNeural2022}
    
    In the literature, there is a strong focus on power-flow related tasks, such as \cite{beinertPowerFlowForecasts2023,gaoPhysicsGuidedGraphConvolution2023,falconerLeveragingPowerGrid2023,liuTopologyAwareGraphNeural2023}, but ML is also increasingly used in the context of dynamical predictions \cite{nandanooriGraphNeuralNetwork2022,sunFastTransientStability2022,zhaoStructureInformedGraphLearning2022,wangNeuralNetworkApplications2023}. Recent studies have demonstrated the efficacy of ML in accurately predicting the dynamical stability of simplified multi-machine systems \cite{nauckPredictingBasinStability2022,nauckDynamicStabilityAnalysis2022,nauckDynamicStabilityAssessment2023,nauckPredictingInstabilityComplex2024}. These studies only consider frequency dynamics for generators and do not properly model line impedances or loads. While these models are too conceptual to allow realistically modeling power grids, these papers show that ML is capable of predicting sophisticated dynamic stability properties.

    Depending on the context, different ML methods have been used. Conventional ML methods, such as gradient boosted trees and deep neural networks, are useful for studying fixed topologies. Graph neural networks (GNNs) are particularly well-suited when topological properties are especially important or when analyzing changing topologies. GNNs can deal with full graphs as inputs, whereas more conventional ML methods consider nodal (respectively bus) features only. Methods that purely rely on nodal features will be referred to as non-graph ML in the following.

    Inspired by the recent success of GNNs, we apply them to analyze dynamical properties of future power grids. In this work specifically, we aim to predict the capabilities of inverter-dominated systems to ride through faults. To tackle this issue, we have overcome several sequential challenges. Firstly, training ML models necessitates extensive datasets. This requires the generation of plausible, synthetic power grid models as detailed in \Cref{sec:dataset_generation}. We generate a substantial dataset that allows us to determine the capabilities of a system to successfully ride-through a fault. Our analysis reveals the significance of the grid topology compared to only studying local bus properties, as shown in section \ref{sec:num_results}. Lastly, in section \ref{sec:ml_results}, we demonstrate the effectiveness of ML models trained on this dataset in predicting the fault ride-through behavior. Furthermore, we show that GNNs trained on synthetic grids exhibit generalization capabilities, predicting the fault ride-through behavior of an IEEE test grid, which is not included in the training dataset.

\section{Theoretical background}
    \subsection{Probabilistic stability analysis} 
    Analyzing power grids under extreme conditions helps to understand the system dynamics and to identify critical components that may fail. Typically, for an evaluated fault scenario a binary classifier is used, meaning that a scenario can either be stable or unstable according to operational bounds or grid codes. While studying extreme scenarios is important, exclusively focusing on a few such scenarios can be limiting \cite{billintonReliabilityAssessmentElectric1994}. When studying a small set of scenarios, the system stability could be overestimated and detrimental situations could be overlooked. To avoid this problem, one can study randomly generated scenarios in a Monte Carlo simulation. 
    
   Such studies are so-called probabilistic stability analyses, that are established for static power flow studies \cite{borkowskaProbabilisticLoadFlow1974} and have recently been mandated by the European Network of Transmission System Operators for Electricity (ENTSO-E) \cite{entso-eAllContinentalEurope2018}. In probabilistic stability analyses a large set of faults is randomly selected to analyze the response of a system. Probabilities of failures, instead of binary classifiers, can then be associated to different grid configurations. 

    Probabilistic approaches are of increasing interest when analyzing the dynamics as well \cite{liemannProbabilisticStabilityAssessment2021, entso-eAllContinentalEurope2018}. They are especially vital in dynamic contexts as power systems are complex, non-linear systems \cite{witthautCollectiveNonlinearDynamics2022}. Identifying the most critical scenarios for non-linear systems is challenging, as a fault with a lower magnitude could lead to a more detrimental system response. A detailed analysis of the non-linear dynamics of the studied dataset shows this effect (\Cref{sec:num_results}). These results clearly underline the need for probabilistic approaches.
        
    The combination of probabilistic analysis with dynamic simulations requires especially expensive computations, which limited their applications in the past. The increasing capabilities of ML-based approaches enable a wider application in the near future.

    While different measures for the dynamic probabilistic stability have been introduced, the focus of this work will be the fault-ride-through probability $p_{frt}$, which will be introduced in \Cref{sec:frt_prob}. 
    
    \subsection{Graph neural networks}
    \label{sec:gnn_background}
    Within the domain of ML, graph neural networks (GNNs) are recognized as the preferred solution for problems that depend on topological properties, as GNNs take the full graph as input. They incorporate the graph via the adjacency matrix or similar representations of the graph connectivity into the ML models. This allows GNNs to aggregate information in the spatial neighborhood. For power grid-related tasks, GNNs can be beneficial when the topology has an important influence, e.g., in case of network congestion, which may result in costly redispatch actions. Conventional ML methods that do not consider topological structure may struggle in those situations. For probabilistic stability, it is known that the topology has a large impact on the behavior \cite{witthautCollectiveNonlinearDynamics2022,nitzbonDecipheringImprintTopology2017}. For GNNs, a breakthrough was the development of the Graph Convolutional Layer (GCN): \cite{kipfSemiSupervisedClassificationGraph2017}:
    \begin{equation}
    	H = \sigma(\overline{A} X \Theta),
    \end{equation}
    where output $H$ is obtained by applying the activation function $\sigma$, using the input features $X$, the learnable parameters $\Theta$ and a normalized and modified version of the adjacency matrix $\overline{A}$. To consider larger surroundings of buses and not only consider direct neighbors, multiple GCN layers can be applied consecutively. Instead, in \cite{duTopologyAdaptiveGraph2017} the topology adaptive graph (TAG) convolution is developed that uses multiple exponents $i$ of $\tilde{A}$ within one layer to consider larger surroundings per layer. The TAG layer can be described by:
    \begin{equation}
    	H = \sum_{z=0}^Z D^{-\frac{1}{2}} A^z D^{-\frac{1}{2}} X \Theta_z.
    \end{equation}
    \subsection{Power grid components}
    Power grids are currently in a transition phase to properly function with increasing shares of renewable energy sources (RES). Variable RES such as solar and wind are connected to the grid via inverters. Currently, most inverters are controlled via grid-following strategies that can not significantly contribute to grid stability when the share of synchronous generation (SG) is low. To secure safe grid operation, transmission operators see an urgency of increasing the share of grid-forming inverters \cite{50hertzNeedDevelopGridforming2020} that contribute to the grid stability in low-inertia systems. Estimates for the required share of grid-forming inverters range from 30 $\%$ \cite{migrateMIGRATENewOptions2019} to 40$\%$ \cite{popellaNecessaryDevelopmentInverterbased2021} of the nominal power of converter interfaced generation in Ireland and Germany, respectively. Future power grids with large shares of RES will behave differently than today's grids. In order to properly study the dynamic stability of inverter-dominated large-scale power grids, models for grid-forming and grid-following inverters are needed. To adequately model the dynamical behavior of grid-forming inverters, the so-called normal form \cite{koglerNormalFormGridForming2022} is employed. The grid-following inverters will be modeled by PQ-buses. 

    \subsubsection{Normal form of grid-forming components}
    There are a variety of proposed grid-forming control strategies, such as dispatchable virtual oscillators \cite{seoDispatchableVirtualOscillator2019} or droop controlled inverters \cite{schifferConditionsStabilityDroopcontrolled2014}.  In future grids, a mix of different grid-forming inverters is to be expected. Furthermore, future grid-forming inverters connected to transmission systems will most likely be black boxes, meaning that their inner control will be unknown, as they are build and sold by private companies. Power grids including grid-forming inverters still have to be simulated adequately to ensure that the large scale integration of grid-forming inverters will not lead to instabilities or undesired interactions between the inverters. It has been shown that different grid-forming models result in very similar transient behaviors \cite{qoriaWP3ControlOperation2018} which indicates that grid-forming behavior can be represented by a uniform technology independent model. 
    
    Kogler et al. \cite{koglerNormalFormGridForming2022} introduced the \textit{normal form} which is a technology-agnostic dynamical model of grid-forming actors. The parameters of the normal form can be directly retrieved from data so that the underlying model does not need to be known, yet simulation studies are still possible. The normal form has been validated by simulations and lab measurements \cite{koglerNormalFormGridForming2022}.
    
    A normal form in the co-rotating reference frame at bus $m$ with a single internal variable, the frequency $\omega_m$, is given by:
    \begin{align}
        \nu_m &= v_m v_m^* \nonumber \\
       \dot{\omega}_m &= B^{\omega, m} \delta \omega_m + C^{\omega, m} \delta \nu_{m} + G^{\omega, m} \delta P_m + H^{\omega, m} \delta Q_m \label{eq:normalform} \\
        \frac{\dot{v}_m}{v_m} &= B^{v,m} \delta \omega_m + C^{v,m} \delta \nu_{m} + G^{v,m} \delta P_m + H^{v,m} \delta Q_k
        \nonumber
    \end{align}
    where $v_m$ is the complex voltage. $\delta P_m$ and $\delta Q_m$ represent the difference between the active and reactive power to the set points $P_{set,m}$ and $Q_{set,m}$. $\delta \nu_m$ is the difference of the squared voltage magnitude $\nu_m$ to the squared nominal voltage. The other coefficients ($B$, $C$ and $G$) are the modeling parameters that capture all the differences between the various models the normal form can represent. In the normal form, all structural differences between models are absorbed in the parametrization.
    
    \section{Fault-Ride-Through Probability}
    \label{sec:frt_prob}
    Fault-ride-through capabilities refer to the ability of a grid-component to stay within a ride-through curve after a fault at a bus has been cleared. They are considered as an important measure of stability, especially for inverters, in various contexts. Examples are research projects focusing on the operation of power grids under 100$\%$ inverters \cite{qoriaWP3ControlOperation2018}, guidelines for grid forming behavior \cite{vdeguidelines_gridforming_2020} or installation rules for power generating facilities \cite{vdeguidelines_connection_2019}. Fault-ride-through capabilities are typically considered as a binary property, given a selected set of faults.
    
    The fault-ride-through probability can be understood as an extension to this concept. Instead of studying a small-set of faults, fault-ride-through statistics are analyzed. In the following, we will refer to the grid state after the occurrence and clearance of a fault as post-clearance states. The fault-ride-through probability $p_{frt}$ is defined by:        
    \begin{align}
        p_{frt} = \frac{V^*}{V_{t}},
    \end{align}
    where $V^*$ is the number of randomly generated post-clearance states for which the transients stay within a ride-through curve and $V_{t}$ is the total number of investigated states. Hence, a system that never exceeds a ride-through curve has $p_{frt}=1$. In this work we consider post-clearance states for individual buses resulting in a fault-ride-through probability for each bus. This allows us to predict critical components that have a high probability of failing after fault clearing. In section \ref{sec:post-clearance-states}, we will provide the methodology employed to randomly generate the set of post-clearance states.
    
    The fault-ride-through probability is a special case of the so-called survivability \cite{hellmannSurvivabilityDeterministicDynamical2016}, an established concept in the theoretical physics community. 
    \subsection{Post-clearance states}
    \label{sec:post-clearance-states}
    To compute the fault-ride-through probability, a large number of post-clearance states per bus is needed. Explicitly modeling every fault to generate these states requires a significant additional modeling effort and computational resources. Instead, so-called Sobol sequences, which are enhanced quasi Monte-Carlo samples \cite{sobolDistributionPointsCube1967} and the ambient forcing algorithm \cite{buttnerAmbientForcingSampling2022} have been employed.
    
    As a first step, the Sobol sequences, which are quasi-random and well-distributed points sequences, are used to sample the space of post-clearance states. The magnitude of the voltage at one of the buses is sampled in a regime of $(0,1)$ \si{p.u.}. Furthermore, the phase angle of the bus is drawn from the regime of (0, +2$\pi$). For grid-forming components, an additional frequency fault in an interval of $(-1, +1)$ \si{Hz} is applied. In a second step, the ambient forcing algorithm is employed as an initialization technique to ensure all algebraic constraints are fullfilled.

    \subsection{Ride-through curves}
    \label{sec:ride_through}
    In this work, we will consider ride-through curves for the bus-voltages and frequencies. For low voltages, we use the same ride-through curve as outlined in \cite{liemannProbabilisticStabilityAssessment2021}. In the case of over-voltages the ride-through curve proposed in \cite{vdeguidelines_connection_2019} has been employed. The frequency ride-through curve is given by constant symmetric thresholds for the frequency of [-2, +2] \si{Hz}. Although we consider post-clearance states for individual buses the fault-ride though curves are always considered for all buses. This means that the fault-ride-through probability is a measure that explicitly considers the dynamics of the entire grid. 


\section{Dataset generation}
\label{sec:dataset_generation}
    Generally, ML methods require sufficiently large datasets to achieve accurate predictions. The computational costs of probabilistic stability assessments scales at least quadratically with the number of buses. The quadratic factor is caused by more expensive computations per fault scenario and the increasing number of faults that have to be analyzed. Hence, calculating $p_{frt}$ of real sized power grids with thousands of buses will remain unfeasible in the near future. Instead of using single real sized power grids that are sufficiently small, we rely on numerous synthetic power grids. By considering numerous different power grids, the training dataset contains many of the topological structures that also exist in real power grids. The following sections describe the process of the dataset generation, including the computationally expensive dynamic simulations.

\subsection{Synthetic power grids} 
    \label{sec:synthetic_grids}
\begin{figure}[H]
        \centering
        \includegraphics[width = \linewidth]{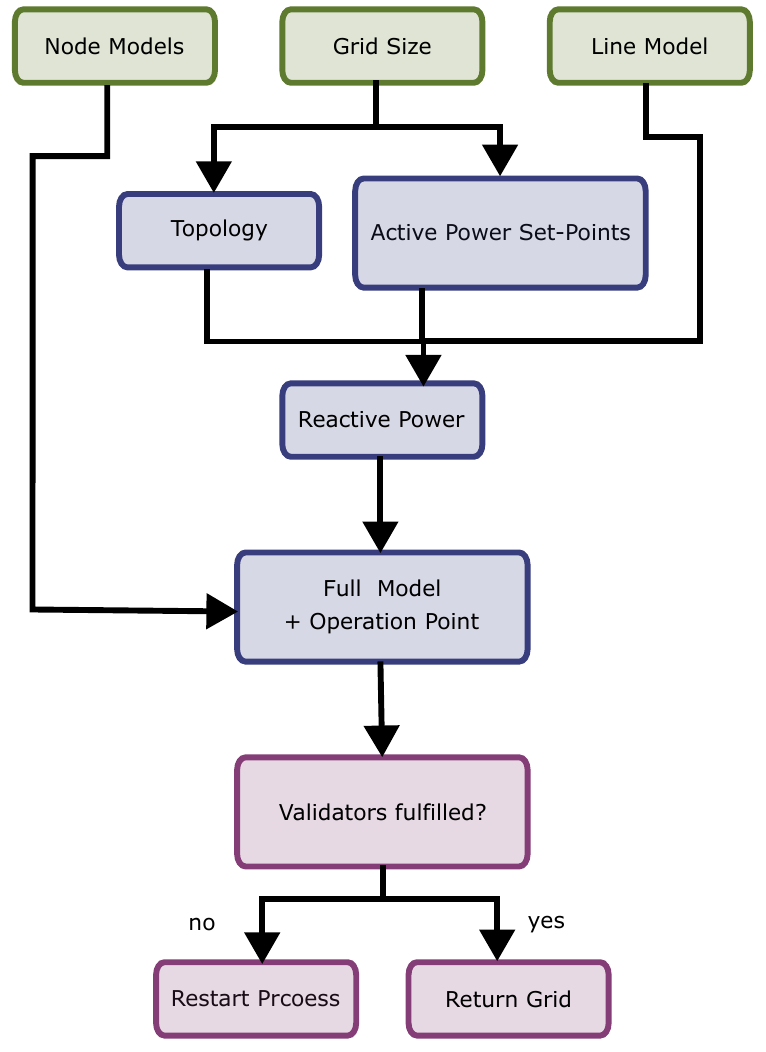}
        \caption{The structure of the power system generation algorithm shows the generation based on the chosen \textcolor{load}{input parameters}. Before the power grid is returned by the software, its behavior is validated to fulfill the stability criteria of real power grids. }
        \label{fig:flow_chart}
    \end{figure}

    In order to generate the dataset, the synthetic power grids framework is used, which has recently been introduced in \cite{buttnerFrameworkSyntheticPower2023}. To keep this work self-contained, a summary of the main features of this algorithm is given. In this work, power grids of different sizes are generated. The number of buses varies between 70 and 80 which is sufficiently large to include topological effects, but is computationally still feasible. The default structure of the algorithm introduced in \cite{buttnerFrameworkSyntheticPower2023} is depicted in \Cref{fig:flow_chart}. Inverter-based power grids with a share of 100$\%$ RES are modeled using the assumption that the grid consists of actors with and without grid-forming capabilities. To model grid-forming and -following components, the normal form \Cref{eq:normalform} and PQ-bus are used respectively. The Pi-model is used to model the transmission lines.
    All following calculations are performed in a Per-Unit-System (p.u.) with a base voltage of \SI{380}{kV} and a base power of \SI{100}{MW}.
    
    The power grid structures are generated by the random growth algorithm that has been introduced in \cite{schultzRandomGrowthModel2014}. The line admittances and shunt admittances are calculated according to the standard line parameters for transmission systems with a voltage of \SI{380}{kV} published by the German energy agency (dena) \cite{denaDenaVerteilnetzstudieAusbauUnd2012}. The parameters are given in \cref{tab:NFparameters}.
    
    For a realistic distribution of active power demand and supply the \emph{ELMOD-DE} dataset \cite{egererDIWBerlinOpen2016}, a dispatch model for the German EHV transmission system, is used. The authors of \cite{taherEnhancingPowerGrid2019} have analyzed the data and suggest a bimodal distribution for the net power $P_{net}$ at a bus. This means that the difference between the produced and consumed power at a bus is taken into account via $P_{net} = P_{prod} - P_{cons}$. The bimodal distribution used to sample $P_{net}$ is given by:
    \begin{align}
        p(P_{net}) = \frac{1}{2 \sigma \sqrt{2 \pi}} \left( \exp{\frac{(P_{net}-P_0)^2}{2\sigma^2}} + \exp{\frac{(P_{net}+P_0)^2}{2\sigma^2}} \right). \label{eq:power_dist}
    \end{align}
    following \cite{buttnerFrameworkSyntheticPower2023} $P_0 = \SI{1.31}{p.u.}.$ is used.
    
    In the introduced framework \cite{buttnerFrameworkSyntheticPower2023} the reactive power is distributed by using a voltage stability objective \cite{zhangReviewReactivePower2007}. The reactive powers at the buses are adjusted to fulfill the desired voltage magnitudes by solving a load flow problem. 
    
    Before a power grid is analyzed further, it is verified that the operation point voltage magnitudes are close to $V \approx 1 \ p.u.$. Furthermore, a small signal stability analysis \cite{milanoPowerSystemModelling2010} is performed to guarantee that the grids are linearly stable. Lastly, it is validated that no lines in the grid are overloaded during normal operation. The stability margin approach, see for example \cite{gonenElectricalPowerTransmission2014}, is used to assure this. 

\subsection{Actor parameterization}
    To generate the datasets, the introduced models for the buses need to be parametrized. The lines have already been parametrized according to section \ref{sec:synthetic_grids}. Bus and line parameters are summarized in table \ref{tab:overhead_lines_parameters}.
    For the normal form, three sets of parameters are used that are published in \cite{koglerNormalFormGridForming2022, buttnerAmbientForcingSampling2022, schifferConditionsStabilityDroopcontrolled2014}. The parametrizations differ in the real parts of $B_x$ and $G_x$ as a consequence of different virtual inertia constants of the grid forming inverters. Since $\mathcal{R} (B_x)$ and $\mathcal{R}(G_x)$ are linearly dependent, we only consider $\mathcal{R} (B_x)$ as a feature for the ML algorithm.  \Cref{tab:NFparameters} summarizes the used parameters. NF1 has the smallest virtual inertia, whereas NF2 and NF3 have increased virtual inertia. Buses with higher inertia are expected to be dynamically more stable.
       
    \begin{table}[h]
        \centering
        \caption{The used parameters for the normal form}
        \begin{tabularx}{\linewidth}{ccccX}
            \toprule
             name & $\mathcal{R}  (B_x)$ &  $\mathcal{R}  (G_x)$ &virtual inertia & source \\
             \midrule
             NF1& -1 & -5& small & Kogler et al. \cite{koglerNormalFormGridForming2022} \\
             NF2 & -2 & -10& medium & Schiffer et al. \cite{schifferConditionsStabilityDroopcontrolled2014}  \\
             NF3 & -0.2& -1& large & Büttner et al. \cite{buttnerAmbientForcingSampling2022} \\
             \bottomrule
        \end{tabularx}
        \label{tab:NFparameters}    
    \end{table}
    
    \begin{table}[H]
        \centering
        \caption{Standard overhead line parameters according to \cite{denaDenaVerteilnetzstudieAusbauUnd2012} for the typical number of cables and wires.}
        \begin{tabularx}{\linewidth}{llll}
            \hline
            \toprule
            Voltage level & $R$ [\si{\Omega \per km}] & $X$ [\si{\Omega \per km}] & $C_{sh}$ [\si{nF \per km}]  \\
            \midrule
            \SI{380}{kV}         &         0.025            & 0.25                    & 13.7  \\ 
            \bottomrule
        \end{tabularx}
        \label{tab:overhead_lines_parameters}
    \end{table}
    
\subsection{Simulation setup}
    When generating the dataset and computing the fault-ride-through probability, there is a compromise between low statistical and numerical errors as well as the computational costs. 1 000 grids with 70 to 80 buses are generated. 1 000 post-clearance state per bus are analyzed and higher order implicit differential algebraic equations solvers are used with low error tolerances, yielding standard errors of $\pm0.02$ for the probabilistic measures. The total computation cost for generating the dataset are roughly 50 000 CPU hours. Details on the software used to generate the dataset can be found in appendix \ref{app:software_dynamics}. The results of the numerical study can be found in \Cref{sec:num_results}.
    

\subsection{IEEE 96-RTS test system}
    The IEEE 96-RTS test case \cite{griggIEEEReliabilityTest1999} was chosen as a test system to validate if the ML models are able to generalize to a different power grid model. The IEEE test system is not used for training, hence the performance is evaluated on a configuration that is entirely new to the ML models. 
    
    The IEEE test case has been updated to also include converter interfaced generation, which is not present in the original system. All buses which generate power are assumed to have grid-forming capabilities and are therefor modeled by the normal form, the same parameters as in \ref{tab:NFparameters} are used as well. The loads, modeled by PQ-buses in the original system, are assumed to have no grid-forming capabilities and thus the model has not been changed in this regard. 
    
    Furthermore, the line parameters of the system are updated to model the same type of transmission system as the synthetic grids introduced in section \ref{sec:synthetic_grids}. The line lengths given in the original test case are kept and the admittances and shunt admittances are calculated according to the dena model \cite{denaDenaVerteilnetzstudieAusbauUnd2012} also employed in the dataset of synthetic power grids. The topology and active power set-points of the test system have not been modified. Finally, the probabilistic dynamic stability of the test system is calculated in the same exact procedure as for the synthetic grids.
    
\section{Machine Learning setup}
    \subsection{Training setup}
\begin{figure}
            \centering
            \includegraphics[width=\linewidth]{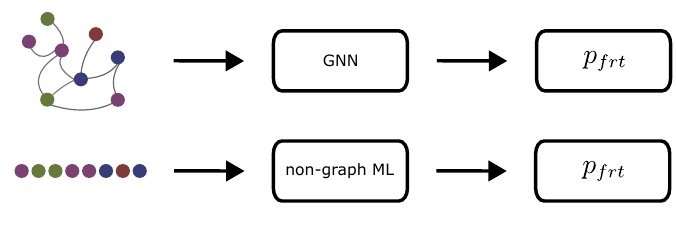}
            \caption{Training setup: The GNN gets the full power grid as input (bus features, line features and the topology) whereas the nodal features are the only input for the non-graph ML methods. In all cases, the outputs are the nodal fault-ride-through probability ($p_{frt}$).}
            \label{fig_training_setup}
        \end{figure}

    To evaluate the capabilities of ML to predict $p_{frt}$ via nodal regression, we compare GNNs (cf.~\Cref{sec:gnn_background}) with non-graph ML predictions using linear regression (linreg) and gradient boosted trees (GBT). By non-graph ML, we mean ML methods, that do not take the topology into account, but purely rely on nodal features. The two general approaches are visualized in \Cref{fig_training_setup}.
    As a loss function, the mean squared error is used and evaluated at all buses except for the slack bus. The next subsection describes the preprocessing of the dynamical results for the ML predictions.

    \subsection{Feature preparation}
    To prepare the information of the dynamical simulations, we need to distinguish the bus types. All three bus types have common nodal features ($P,Q$) and the Normal Form buses have one additional relevant input parameter: $\mathcal{R} (B_x)$. 
    
    To improve the predictive power of all ML methods, we compute three additional nodal features based on the surrounding line properties of each bus $k$. The three used power lines parameters are: the conductance $G_{km}$, susceptance $B_{km}$ and the shunt susceptance $B_{km, sh}$. Computing the sums yields: $G_k = \sum_m G_{km}$, $B_k = \sum_m B_{km}$ and $B_{k, sh} = \sum_m B_{km, sh}$. Furthermore, two additional categorical parameters are used as input for the bus category to describe if a bus is a slack or a load. To have the same number of input parameters for all bus types, 0 is used for $\mathcal{R} (B_x)$ for load and slack buses. This leads to 8 input features in total that are directly used for GBT and linear regression as only input and in case of the GNNs as nodal features. 
        
    Further, the input features may be scaled as a common preprocessing step. In case of GNNs and linear regression, standardization is used, whereas no scaling is used for GBT. 

\subsection{Performance measures}
    To evaluate the performance of the ML models, two measures are used. First, the coefficient of determination $R^2$  represents the proportion of the variation of the data explained by the model. A perfect model would have $R^2 = 1$, where a model that predicts the average of the test distribution achieves $R^2 = 0$. Second, Spearman's rank correlation coefficient $\rho$ is used, to analyze if the model predicts the correct ranking. For real world applications, the identification of critical components might be more important than predicting the correct value. In this case, ML can be thought of as a vulnerability detection tool. Based on the ranking of ML methods, a more detail analysis for the predicted critical component can be carried out.

\section{Results}
\subsection{Numerical results of dynamical simulations}
    \label{sec:num_results}
\begin{figure}[]
        \centering
        \includegraphics[width=.85\linewidth]{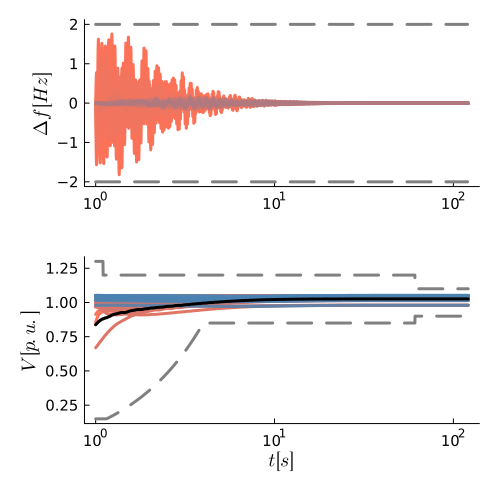}
        \hfil
        \includegraphics[width=.85\linewidth]{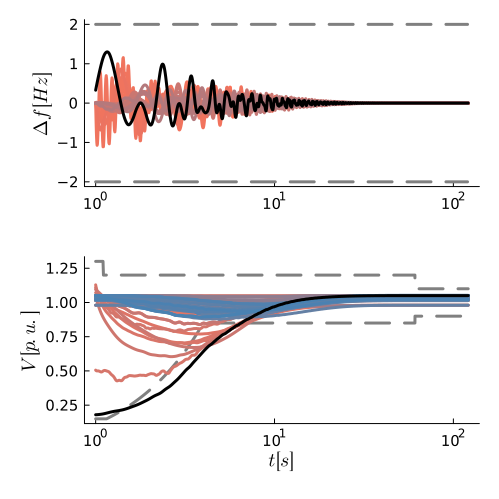}
        \caption{Example transients of the dynamical simulations, visualized by the frequency deviation from the reference frequency of 50 Hz and the voltage magnitude $V$. The plots visualize a stable configuration (top) and an unstable configuration (bottom), for which the voltage thresholds are exceeded. The dashed line in \textcolor{traj_gray}{gray} denotes the operational boundaries. The black transients mark the bus where the fault occurred. The time is plotted on a logarithmic scale to highlight the first time steps.}
    \label{fig::sim_trajectories}
    \end{figure}

    First, the results of individual transients are introduced, before considering multiple scenarios and studying the statistical properties of the datasets. Examples of individual transients of the dynamic simulations are shown in \Cref{fig::sim_trajectories}. The binary classifiers for stability, given by the fault ride-through-curves introduced in section \ref{sec:ride_through}, are evaluated for each simulation. The two cases visualize a stable and an unstable transient. The unstable transient exceeds the voltage ride-through limit. 

\begin{figure}
    \centering
        \includegraphics[height=0.9\textheight]{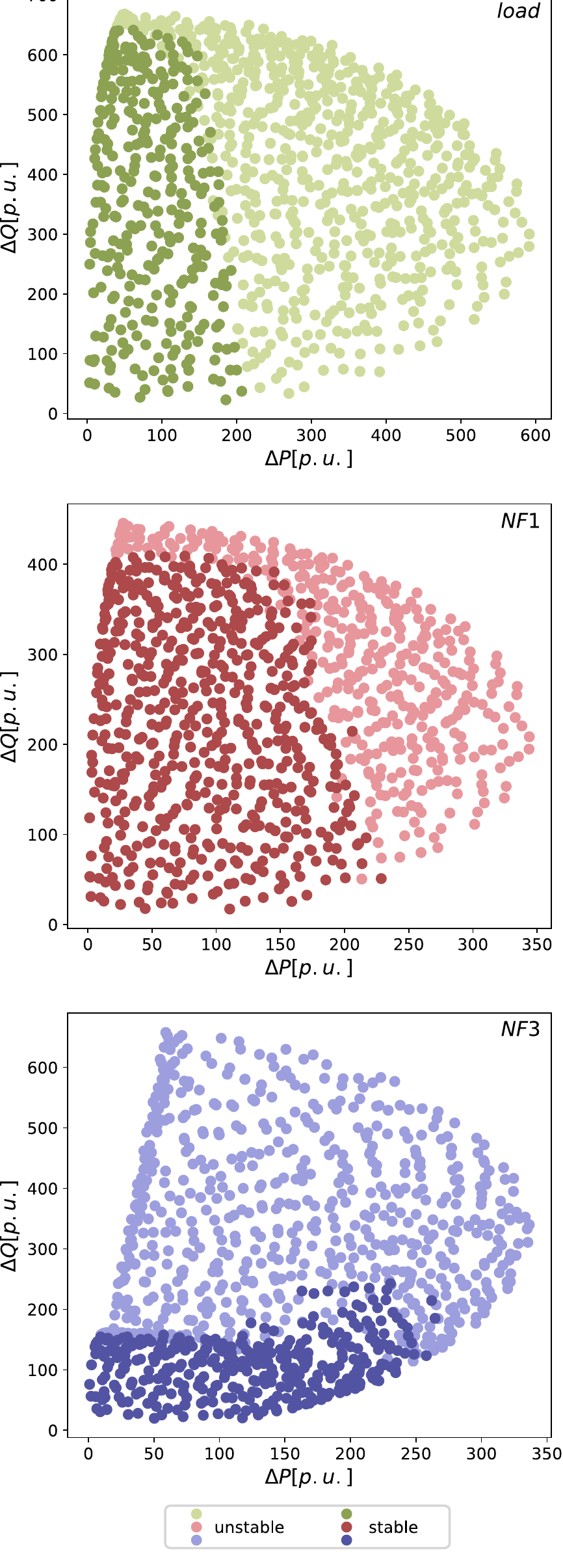}
        \caption{The active and reactive power deviations $\Delta P$ and $\Delta Q$ resulting from the faults at three different buses of the IEEE test case, and the outcome of the corresponding simulations. Darker points indicate that the systems survives and returns to stable operation, whereas light points indicate a failure. Color distinguishes between the types of the shown buses, namely a load bus (top), a NF1 bus (mid), and a NF3 bus (bottom).}
    \label{fig:pert-landscape}
    \end{figure}

    We further analyze the dynamical behavior by studying exemplary outcomes for three different bus types in \Cref{fig:pert-landscape}. Each fault results in a system-wide mismatch in active $\Delta P$ and reactive power $\Delta Q$ in the post-clearance states. The figure plots the post-clearance states in the $\Delta P$ and $\Delta Q$ plane. Solid markers and light markers represent post-clearance states that result in a successful and unsuccessful fault-ride-through, respectively.

    In many cases, although the power differences are similar, the scenarios result in different stability outcomes. Importantly, the figures illustrate cases where larger magnitudes of power differences lead to stable outcomes, whereas smaller differences lead to instabilities. The behavior is thus clearly non-linear. This shows that the outcome of individual fault scenarios are difficult to predict by solely looking at the magnitude of the fault. Dangerous situations might be missed when only analyzing a small number of severe faults. This highlights the need for probabilistic stability analyses to uncover critical components in the grid that may fail in a wide range of situations.

\begin{figure}[t]
    \centering
        \includegraphics[width=\linewidth]{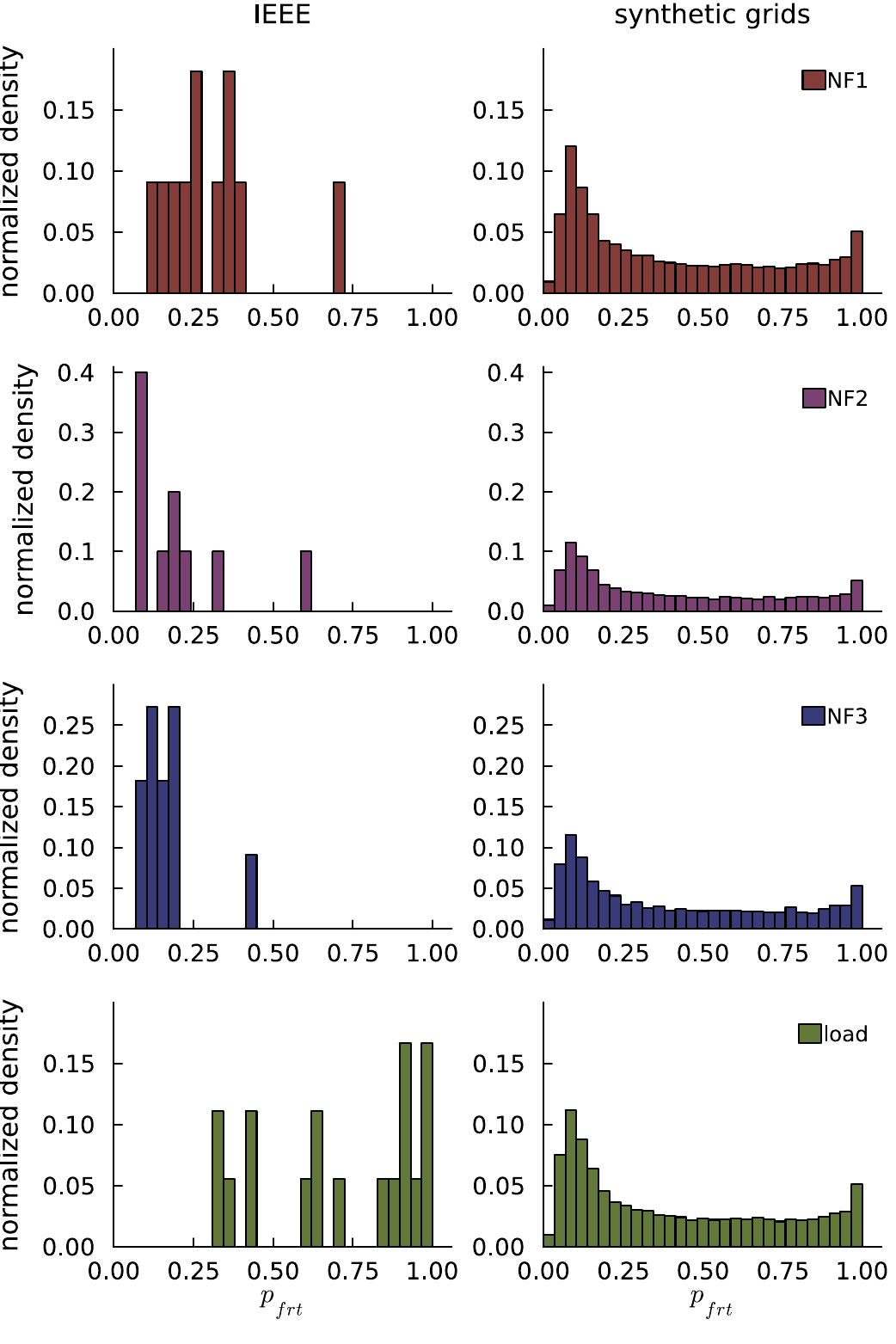}
        \vspace{-11pt}
        \caption{The histograms of the $p_{frt}$ for the IEEE test case are on the left and for the 1 000 synthetic grids on the right. The rows depict the four different bus types. NF1-3 denotes the different normal form parameterizations. For the loads, a different scaling of the vertical axis is used.}
    	\label{fig_HistogramsSurv}
    \end{figure}

    Finally, we discuss the statistical properties of the dataset. The histograms of $p_{frt}$ are shown in \Cref{fig_HistogramsSurv}. For the synthetic grids, we see that different virtual inertia constants lead to different fault-ride-through behaviors. As expected, the buses with the highest virtual inertia, $NF3$, lead to the highest mean fault-ride-through probability. In all histograms, we see that the $p_{frt}$ varies vastly for individual buses. Buses with the same machine parameters can either always successfully ride-though a fault or might never, or anything in between. We clearly see an impact of the grid topologies on the stability which can not be explained on the individual component level. Hence, it is not sufficient to only study individual components, but the topological embedding has to be considered as well. 

    The IEEE test case is slightly more stable ($\overline{p_{frt}}$ $\approx .589$) in comparison to the synthetic power grids ($\overline{p_{frt}} \approx .522$). In general, such distribution shifts are challenging for the generalization capabilities of ML approaches, because ML models are good at interpolating, but struggle at extrapolating to unseen data. Hence, the task of predicting the dynamic stability of the IEEE test case can be seen as a test for the generalization capabilities of the used ML models. Strong generalization capabilities indicate that the models do not only learn training-set specific patterns, but learn more causal, which is promising for a wider range of applications.

    We see that our simulations feature both, non-linear dependence of the stability outcomes on the fault magnitudes, and a clear impact of the topology on $p_{frt}$. These results further motivate the probabilistic approach for determining stability. Because of the impact of the topology on the stability, we suspect that GNNs will have an advantage over the non-graph ML methods.
    
    \subsection{ML performance}
    \label{sec:ml_results}
    The results in \Cref{tab:results} show that ML can be utilized to predict $p_{rtf}$. All ML methods are able to predict $p_{rtf}$ on the synthetic power grids, and they also generalize to the IEEE test case. Importantly, no strong statistical conclusions can be drawn from a single IEEE test case as it only consists of 72 buses respectively data points. The IEEE test case should be considered as an illustration of strong generalization capabilities.
    
\begin{figure}
        \centering
        \vspace{-.1cm}
        \includegraphics[width=\linewidth]{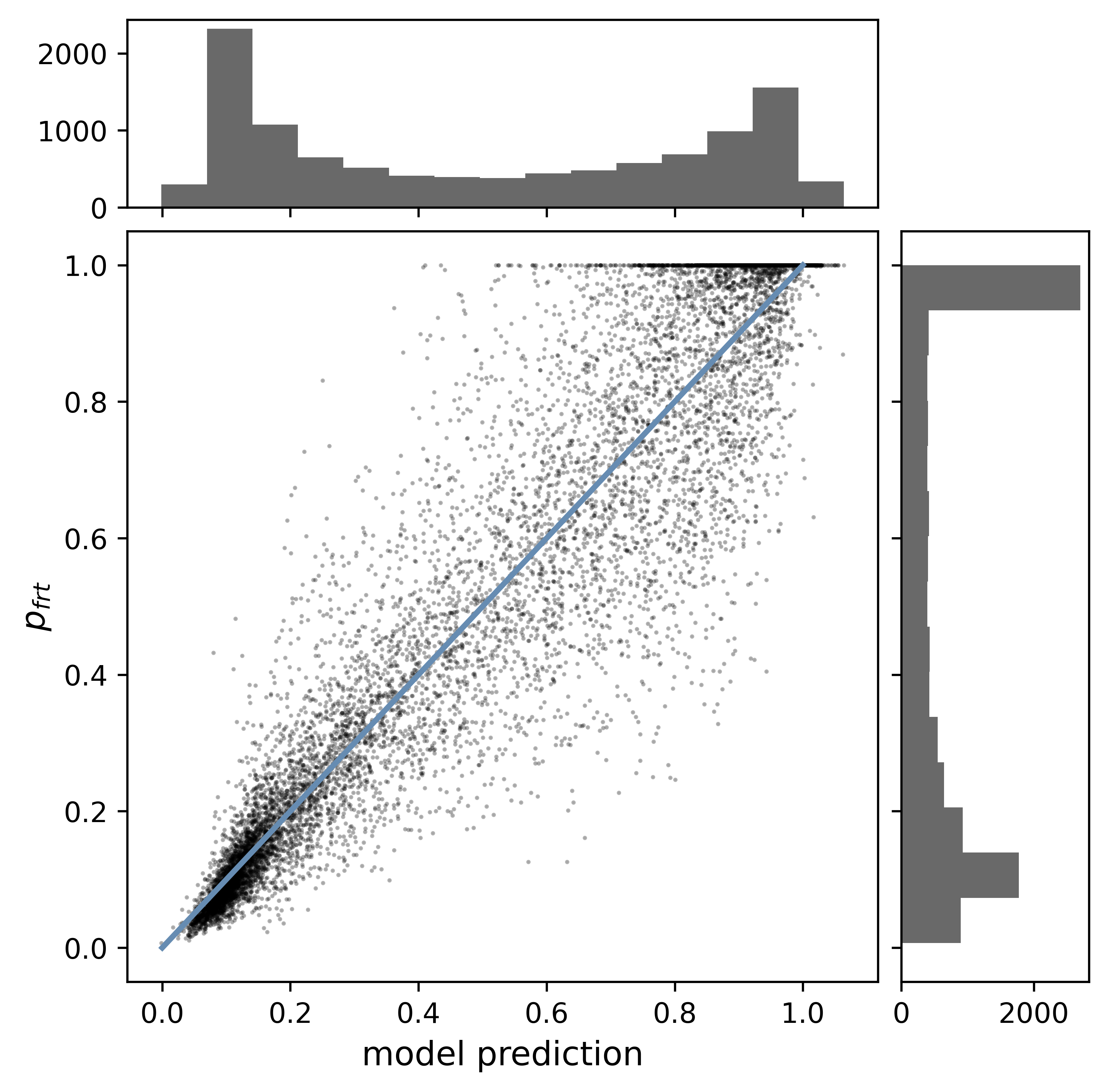} 
        \caption{Visualization of the performance of TAG. A perfect model would be on the diagonal \textcolor{blue_line}{blue} line.} 
        \label{fig:results_scatter_both}
        \vspace{-.2cm}
    \end{figure}

    On the synthetic power grids, TAG clearly outperforms non-graph ML (regression and GBT). \Cref{fig:results_scatter_both} visualizes the performance of the TAG model. The majority of the predictions are around the diagonal line, indicating a high predictive performance. Considering the histograms of the labels (true $p_{frt}$) it can be seen that there are two peaks in the distributions representing less stable configurations ($\approx .2$) and more stable buses ($\approx .9)$. Both peaks can be predicted correctly, as can be seen in the black regions indicating many predictions. 
    
    Even though TAG achieves better performance on the synthetic datasets, GBT achieve a better performance on the IEEE test case. Apparently, the TAG does not generalize well to the IEEE test case. However, by increasing the regularization of the GNN using dropout (see \Cref{sec_regularization}), denoted by the model TAGreg, the performance can be increased on the IEEE test case, which is bought by a lower performance on the synthetic datasets. The regularized GNN (TAGreg) outperforms GBT on all tasks. This is to be expected as the TAGreg includes topological features which have an impact on the $p_{frt}$ as shown in section \ref{sec:num_results}.
    In general, the solid performance of all methods show that ML is able to predict the dynamic stability of the IEEE case, even though its properties are quite different from the synthetic power grids. 
    
        \begin{table}[h]
            \centering
            \small
            \caption{Results of predicting survivability using $R^2$ and $\rho$ in \% with different ML applications. TAGreg denotes a TAG-model with enhanced regularization.}
            \begin{tabularx}{\linewidth}{Xrrrr}
                \toprule
                 model & test $R^2$ & ieee  $R^2$ & test $\rho$ & ieee $\rho$\\
                 \midrule
                 linreg  & 62.11 & 72.02 & 74.56 & 81.86\\
                 GBT & 62.99 & 72.17 & 74.81 & 76.13\\
                 TAG &   \textbf{90.36} \tiny{$\pm 0.15$} &  60.46 \tiny{$\pm 5.59$}&  \textbf{95.05} \tiny{$\pm 0.07$}& 83.37 \tiny{$\pm 1.79 $}\\
                 TAGreg &  71.51 \tiny{$\pm 1.14$} &  \textbf{77.47} \tiny{$\pm 0.13$}&  86.12 \tiny{$\pm 0.06$}& \textbf{87.32} \tiny{$\pm 0.37 $}\\
                 \bottomrule
            \end{tabularx}
            \label{tab:results}
        \end{table}
    
\section{Conclusion}

    Conducting and analyzing large sets of fault scenarios is required by ENTSO-E \cite{entso-eAllContinentalEurope2018} to assess the stability of power grids. The feasibility of these approaches has been limited in the past due to high computation times of the dynamical simulations. In our work, we study the potential of using ML methods to predict the outcome of such probabilistic stability assessments. We especially focus on the fault-ride-through probability ($p_{frt}$) of power grids with large shares of inverters.  
    
    To train the ML model, we generated a new dataset of inverter-based power grids. The dataset highlights both, the relevance of topological structures, as well as the non-linear character of power-system stability.
    
    All deployed ML methods are able to predict $p_{frt}$. The Graph Neural Networks (GNNs), which consider topological features, achieve a better performance than non-graph-based ML methods. Importantly, ML methods can generalize to a differently structured IEEE test case, that was not part of the training set. This highlights the potential of training ML methods on synthetic datasets and applying them to real-sized power grids, for which probabilistic stability assessments are not computationally feasible. 
    
    To conclude, the study shows the potential of applying ML to predict probabilistic stability measures of future power grids. Due to the fast evaluation times, there are almost no limits for the number of configurations that can be analyzed using a fully trained model. This can help to improve the quality of power grid operation by enabling the automatic analysis of numerous configurations and identification of critical components that are prone to fail. Furthermore, the application of ML could aid by considering dynamic effects at early planning stages, leading to optimal system designs that avoid costly dynamical stabilization measures.


\section*{Acknowledgments}
    All authors gratefully acknowledge Land Brandenburg for supporting this project by providing resources on the high-performance computer system at the Potsdam Institute for Climate Impact Research. The work was in parts supported by DFG Grant Numbers KU 837/39-2 (360460668), BMWK Grant 03EI1016A and BMBF Grant 03SF0766. Michael Lindner greatly acknowledges support by the Berlin International Graduate School in Model and Simulation (BIMoS) and by his doctoral supervisor Professor Eckehard Schöll. Christian Nauck would like to thank the German Federal Environmental Foundation (DBU) for funding his PhD scholarship and Professor Jörg Raisch for the supervision. Anna Büttner acknowledges support from the German Academic Scholarship Foundation and her doctoral supervisor Professor Jürgen Kurths. AI tools are used on the (sub)-sentence level to improve language.

\appendix

\subsection{Data and source code availability}
    For reviewing and upon publication, the code to generate the datasets and figures, as well as to train the ML models is provided on \url{https://github.com/PIK-ICoNe/FaultRideThroughProbabilityML_paper-companion.git} and \url{https://zenodo.org/record/11193718}.


\subsection{Details on generating the datasets}
    \label{app:software_dynamics}
    To generate the datasets, Julia 1.9.1 and the packages PowerDynamics.jl \cite{plietzschPowerDynamicsJlExperimentally2022} and NetworkDynamics.jl \cite{lindnerNetworkDynamicsJlComposing2021}, that heavily rely on DifferentialEquations.jl \cite{rackauckasDifferentialEquationsJlPerformant2017}, were used for solving differential equations.


\subsection{Training details}
    This section includes more information to reproduces the obtained results. For the training, Pytorch \cite{paszkePyTorchImperativeStyle2019} is used. For the graph handling and graph convolutional layers the library PyTorch Geometric \cite{feyFastGraphRepresentation2019} is used. 

\paragraph{Hyperparameter study}
    ML models and their training are characterized by so-called hyperparameters. Hyperparameters are not learnable during the training, but distinctive for the behavior, for example by specifying the size of the models. Hyperparameter studies are conducted to optimize the performance by varying the number of hidden dimensions, scaling strategies, learning rates, dropout and model-specific parameters such as the number of hops in case of TAG ($K$). For the hyperparameter study, ray \cite{moritzRayDistributedFramework2018} and tune \cite{liawTuneResearchPlatform2018} are used. 

\paragraph{Final properties of the model}
    The TAG-model has the following properties: 3 layers, 304 hidden channels, the exponent $K=3$, dropout $\approx 0.35$, and one linear layer with dimension 500 after the final graph convolution and the batch size is 700, and consists of 750 273 parameters in total. The regularized model has dropout $\approx 0.43$.

\subsection{Regularization methods for ML}
\label{sec_regularization}
One of the common problems of ML is the tendency of overfitting. By overfitting, we mean a process during training, when the performances on the training set increase, whereas the validation and test performances do not increase anymore. To avoid models from simply memorizing the training data, dropout \cite{hintonImprovingNeuralNetworks2012} is a widely used regularization method. The concept of dropout is applied during training, when a pre-defined ratio of neurons is randomly deactivated.
\printbibliography

\newpage

\end{document}